\pdfoutput=1

\IfFileExists{latexml.sty}{\RequirePackage{latexml}}{\newif\iflatexml\latexmlfalse}

\iflatexml 
\else 
\fi 

\documentclass[conference]{IEEEtran}  \newcommand \equalizelastpgcols  {0mm}

\usepackage[utf8]{inputenc}
\usepackage[T1]{fontenc}

\def\equalizetitlepagecols {0mm} 
\IEEEoverridecommandlockouts

\usepackage{amsmath,amsfonts} \usepackage{array}
\iflatexml 
\else 
\usepackage{tgcursor}
\usepackage{algorithm} 
\fi 

\usepackage{comment}
\usepackage{alltt} 
\usepackage{upquote} 

\usepackage{graphicx} 

\usepackage{amsthm} 

\newlength{\bufferhline} 

\def\myblue{blue}
\def\myblack{black}

\iflatexml 
\usepackage{color}
\usepackage{soul}
\setulcolor{blue}
\usepackage[colorlinks=true]{hyperref}
\usepackage[square,numbers]{natbib}
\newcommand{\shortdoi}[1]{\href{https://doi.org/#1}{\ul{\color{\myblack}\texttt{doi:#1}}}}
\newcommand{\wvcite}[1]{\citep[{\color{\myblue}{$\Rightarrow$}}][]{#1}}
\newcommand{\wvref}[1]{\hyperref[#1]{{\color{\myblue}{$\Rightarrow$}}{\color{\myblack}{\ref{#1}}}}}
\newcommand{\wveqref}[1]{\hyperref[#1]{{\color{\myblue}{$\Rightarrow$}}{\color{\myblack}{\eqref{#1}}}}}
\def\mybegineqstar#1{\begin{equation}\label{#1}}
\def\myendeqstar{\end{equation}}
\def\mytag#1{}
\else 
\usepackage[normalem]{ulem}
\usepackage[svgnames]{xcolor}
\usepackage[colorlinks=true,allcolors=\myblack]{hyperref}
\usepackage[table]{hypcap}
\newcommand{\shortdoi}[1]{{\color{\myblue}\uline{\color{\myblack}\href{https://doi.org/#1}{\texttt{doi:#1}}}}}
\newcommand{\wvcite}[1]{{\color{\myblue}\uline{{\color{\myblack}\cite{#1}}}}}
\newcommand{\wvref}[1]{{\color{\myblue}\uuline{{\color{\myblack}\ref{#1}}}}}
\newcommand{\wveqref}[1]{{\color{\myblue}\uline{{\color{\myblack}\eqref{#1}}}}}
\def\mybegineqstar#1{\hbox{\refstepcounter{equation}\label{#1}}\begin{equation*} }
\def\myendeqstar{\end{equation*}}
\def\mytag#1{\tag{#1}}
\fi 

\newcommand \doichoice [1] {\shortdoi{#1}} 

\iflatexml 
\else 
\usepackage{afterpage}
\usepackage{footnotehyper}
\fi 

\newcommand \nbdd {\nobreakdash}

\newlength{\myrowskip}

\iflatexml 
\newcommand{\myTwoColCaptionvspace} [1][]{}
\newcommand{\myTwoColSectionvspace} [1][]{} 
\newcommand{\myTwoColTableLiftvspace} [1][]{}
\newcommand{\mydepthbox}{\vrule height0pt depth10pt width0pt}

\else 
\newcommand{\myTwoColCaptionvspace} [1][-2.0mm]{\vspace{#1}}
\newcommand{\myTwoColSectionvspace} [1][-2.0mm]{\vspace{#1}} 
\newcommand{\myTwoColTableLiftvspace} [1][-3.0mm]{\vspace{#1}} 
\newcommand{\mydepthbox}{\vrule height0pt depth0pt width0pt}

\fi 

\iflatexml 

\newtheorem{theorem}{Theorem}
\newtheoremstyle{mystyle}%
  {}
  {}
  {\noindent\itshape}
  {}
  {\bfseries\large}
  {~:}
  {\newline}
  {\noindent\thmname{#1}\thmnumber{ #2}\thmnote{ (#3)}} 
\theoremstyle{mystyle}
\newtheorem{theorem}{Theorem}
\newtheorem{algorithm}{Algorithm(s)}

\def\myabstract{myabstractenv}
\def\mykeywords{mykeywordsenv}

\else 

\def\myabstract{abstract}
\def\mykeywords{IEEEkeywords}

\fi 

\def  \BeginAlgFigure {\begin{algorithm}}
\def  \EndAlgFigure {\end{algorithm}}

\usepackage[english]{babel} 

\makeatletter
\let\ORIbbl@fixname\bbl@fixname
\def\bbl@fixname#1{%
  \@ifundefined{languagealias@\expandafter\string#1}
    {\ORIbbl@fixname#1}
    {\edef\languagename{\@nameuse{languagealias@#1}}}%
}
\newcommand{\definelanguagealias}[2]{%
  \@namedef{languagealias@#1}{#2}%
}
\let\BIBforeignlanguage\foreignlanguage
\makeatother

\definelanguagealias{en}{english}

\begin{document}
\iflatexml 
\def\mythankstext{\copyright The author, dunning@bgsu.edu, %
grants copyright permissionsas specified in\\
Creative Commons License CC BY-SA 4.0. (See Endnote 2)\\ 
\\
}
\else 
\def\mythankstext{\copyright The author  grants copyright permissions as specified in %
Creative Commons License CC BY-SA 4.0.  Attributions should reference this manuscript. %
Version of \today.
}
\fi 
\newcommand \thankstext {\mythankstext}

\title{\LARGE A Permutation-Free Length 3 Decimal Check Digit Code \\[-1.2ex]}

\author{\IEEEauthorblockN{Larry A. Dunning%
\iflatexml 
\\[0.5ex]%
\else 
, Prof. Emeritus}%
\fi 
\thanks{\thankstext}
\IEEEauthorblockA{\small Dept. of Computer Science\\
Bowling Green State University\\
Bowling Green, Ohio 43403, USA\\
Email: dunning@bgsu.edu} \\[-3.0ex]} 

\makeatletter
\def\myDoublePARstart{\@IEEESAVECMDIEEEPARstart}
\makeatother

\maketitle

\iflatexml 
\else 
\makeatletter
\let\lastpagenumber2 
\def\ps@plain{%
  \let\@mkboth\@gobbletwo
   \renewcommand{\@oddhead}{}%
  \renewcommand{\@evenhead}{}%
  \renewcommand{\@evenfoot}{\reset@font\small\hfil\thepage{}~of~\lastpagenumber \hfil}%
  \renewcommand{\@oddfoot}{\@evenfoot}%
}
\pagestyle{plain} 
\makeatother
\thispagestyle{plain}
\fi 

\begin{\myabstract}
In 1969 J. Verhoeff provided the first examples of a decimal error detecting code 
using a single check digit to provide protection  against all single, 
transposition and adjacent twin errors.
The three codes he presented are length 3\nbdd-digit codes with 2 information digits.
Existence of a 4\nbdd-digit code would imply the existence of 10 such disjoint 3\nbdd-digit codes. 
Apparently, not even a pair of such disjoint 3\nbdd-digit codes is known.  
The code developed herein, has the property that the knowledge of any two digits is
sufficient to determine the entire codeword even though their positions were unknown.
This fulfills Verhoeff's desire to eliminate ``cyclic errors".
Phonetic errors, where 2 digit pairs of the forms X0 and 1X are interchanged, 
are also eliminated.  
\end{\myabstract}

\begin{\mykeywords}
Decimal error detection, transposition errors, 
twin errors, phonetic errors, cyclic errors, permutation-free.
\end{\mykeywords}

\enlargethispage{\equalizetitlepagecols}

\myTwoColSectionvspace
\section{Verhoeff's 3-digit Decimal Codes}
\iflatexml 
In
\else 
\myDoublePARstart{I}{n} 
\fi 
his 1969 monograph Jacobus Verhoeff \wvcite{Verhoeff1969}
presented 
three variations of the  ``curious 3\nbdd-digit decimal code", derived from a block design.
shown in Table~\wvref{VerhoeffRegular}.
Each table entry gives the middle digit, $s$, of the codeword throughout this paper, 
allowing a simpler correspondence between properties of the table 
and the requirements for detecting various error types.  
An error in just one of the three digits 
is termed a \textit{single error} or \textit{transcription error}. 
Note that the single error detecting property implies that 
knowledge of any two digits and their positions determine the remaining digit.  
Detection of single errors requires 
that each row and each column of the table be a permutation.  
Detection of \textit{twin errors}, 
where two identical adjacent digits change identically, 
requires that these permutations each have at most one fixed point.  
Detection of \textit{transposition errors}, 
where two differing adjacent digits become interchanged, 
requires that these permutations have no cycles of length~2. 
Detection of  \textit{jump twin errors}, 
where a codeword \texttt{(aba)} is interchanged with \texttt{(cbc)}, 
requires that the main diagonal be a permutation.   
Detection of \textit{jump transposition errors}, 
where \texttt{(abc)} is exchanged with \texttt{(cab)}, 
is provided when the table is completely asymmetric about the main diagonal.

All the error types listed above are provided by Verhoeff's three codes.
In addition, the ``Irregular Code" of Table~\wvref{VerhoeffIrregular}, detects phonetic errors,
while the code of Table \wvref{VerhoeffRegular}  has
codewords \texttt{(302)} and \texttt{(132)} yielding a \textit{left phonetic error} 
and codewords \texttt{(230)} and \texttt{(213)} yielding a \textit{right phonetic error}. 
The code developed in the next section handles these error types.
Like Verhoeff's codes, it fails to detect triple errors.

\iflatexml 
\else 
\def\noteone{In the original tables on pages 40-41 of \wvcite{Verhoeff1969}, 
the table ``resulting" from the block design 
is erroneously interchanged with the ``interchanged" table (not shown) in which the codeword ``\texttt{(109)}" appeared as ``\texttt{(100)}".}
\fi 
\begin{table}[ht] \label{VerhoeffRegular}
\iflatexml 
\caption[]{Verhoeff's Block Design Code$\textsuperscript{\wvref{foot1}}\mydepthbox$\\}
\else 
\caption[]{Verhoeff's Block Design Code$\color{blue}\uuline{\protect\footnote{\noteone}}$}
\def\noteone{In the original tables on pages 40-41 of \wvcite{Verhoeff1969}, 
the table ``resulting" from the block design 
is erroneously interchanged with the ``interchanged" table (not shown) in which the codeword ``\texttt{(109)}" appeared as ``\texttt{(100)}".}
\fi 
\centering%
\myTwoColCaptionvspace
\setlength{\tabcolsep}{0.5em}
\begin{tabular}{c|cccccccccc}
$S$&0&1&2&3&4&5&6&7&8&9\\
\hline
\rule{0pt}{2ex}   
0&0&3&1&2&9&4&5&6&7&8\\
1&2&1&3&0&5&8&7&4&9&6\\
2&3&0&2&1&7&6&9&8&5&4\\
3&1&2&0&3&8&9&4&5&6&7\\
4&5&7&9&6&4&1&8&2&3&0\\
5&6&4&8&7&0&5&2&9&1&3\\
6&7&9&5&8&3&0&6&1&4&2\\
7&8&6&4&9&1&3&0&7&2&5\\
8&9&5&7&4&6&2&3&0&8&1\\
9&4&8&6&5&2&7&1&3&0&9\\
\end{tabular}
\end{table}
\begin{table}[ht]\label{VerhoeffIrregular}
\myTwoColTableLiftvspace
\iflatexml 
\caption{Verhoeff's Irregular Code\textsuperscript{\wvref{foot1}}\mydepthbox}
\else 
\caption{Verhoeff's Irregular Code} %
\fi 
\centering
\myTwoColCaptionvspace
\setlength{\tabcolsep}{0.5em}
\begin{tabular}{c|cccccccccc}
$S$&0&1&2&3&4&5&6&7&8&9\\
\hline
\rule{0pt}{2ex}   
0&0&3&4&9&6&7&5&8&2&1\\
1&5&1&0&2&8&3&9&6&7&4\\
2&7&6&2&4&1&0&8&9&3&5\\
3&1&5&8&3&7&6&4&0&9&2\\
4&2&9&7&5&4&8&1&3&0&6\\
5&6&7&9&0&3&5&2&4&1&8\\
6&3&8&1&7&5&9&6&2&4&0\\
7&9&4&5&8&2&1&0&7&6&3\\
8&4&0&6&1&9&2&3&5&8&7\\
9&8&2&3&6&0&4&7&1&5&9\\
\end{tabular}
\myTwoColCaptionvspace
\end{table}
\iflatexml 
\else 
\afterpage{\setcounter{footnote}{1}\footnotetext{In the original tables on pages 40-41 of \wvcite{Verhoeff1969}, 
the table ``resulting" from the block design 
is erroneously interchanged with the ``interchanged" table (not shown) in which the codeword ``\texttt{(109)}" appeared as ``\texttt{(100)}".}}
\fi 
\myTwoColSectionvspace[-6.0mm]
\section{A Permutation-Free 3-digit Decimal Code}
The code shown in Table \wvref{Code47374800} 
provides all the protections cited for the previous codes.  
The code shown also protects against all phonetic errors.  
The main cited advantage of the the ``Irregular Code" of Table~\wvref{VerhoeffIrregular}  was
that it provides protection against 
all but 16 \textit{cyclic errors}, where \texttt{(abc)} is exchanged with \texttt{(bca)},
while the other two codes given in \wvcite{Verhoeff1969} provide no protection. 
The code given in Table~\wvref{Code47374800}, however, detects all cyclic errors.
In fact, any permutation of the digits of a codeword is detected.
\begin{table}[ht]
\myTwoColTableLiftvspace
\caption{A Permutation-Free Decimal Code\mydepthbox}\label{Code47374800}
\centering
\myTwoColCaptionvspace
\setlength{\tabcolsep}{0.5em}
\begin{tabular}{c|cccccccccc}
$S$&0&1&2&3&4&5&6&7&8&9\\
\hline
\rule{0pt}{2ex}   
0&0& 9& 7& 1& 2& 3& 4& 8& 6& 5\\
1&5& 1& 3& 7& 6& 9& 2& 0& 4& 8\\
2&8& 4& 2& 6& 3& 0& 7& 5& 9& 1\\
3&2& 5& 8& 3& 7& 4& 9& 6& 1& 0\\
4&1& 7& 6& 0& 4& 2& 8& 9& 5& 3\\
5&6& 8& 1& 9& 0& 5& 3& 4& 7& 2\\
6&9& 3& 0& 8& 5& 7& 6& 1& 2& 4\\
7&4& 2& 9& 5& 8& 1& 0& 7& 3& 6\\
8&3& 0& 5& 4& 9& 6& 1& 2& 8& 7\\
9&7& 6& 4& 2& 1& 8& 5& 3& 0& 9\\
\end{tabular}
\end{table}

\section{Construction of the Permutation-Free Code}\label{construction}
Following McKay et al. \wvcite{McKay2007}, the latin square of Table~\wvref{Code47374800}
was represented as an orthogonal array using a set $L$ of ordered ``triplets", with rearrangement to
place the symbol $s$ in the middle. The integer set/sequence
$\{a, a{+}1, \ldots, b\}$ will be denoted by $I_a^b$ or simply $I_n$ for $I_0^{n-1}$and the cardinality of a set $A$ by $|A|$.
The symbols $\forall$ and $\ni$ are used to abbreviate ``for all" and ``such that."
The entries of $L = N \cup D$ are divided into non-diagonal $N$
and diagonal $D$ sets of entries.
\begin{equation*}
\begin{aligned}
	&N=\{ (r_i  s_i  c_i) \;\ni\; | \{r_i  s_i  c_i\}| = 3 \land i \in I_{90}  \} \\
	&D = \{(i \,i \,i) \;\ni\; i {\in}I_{10} = \{0, \ldots, 9\} \}
\end{aligned}
\end{equation*}
There are $720=10*9*8$ possible entries for the $90$ positions in $N$
to provide a guarantee that all codewords 
that may be retained in $N$ will be composed of three distinct symbols.
Let $C = \{\, C_i \;|\; i\in\,I_{120} \}$ represent the collection of the 
$120$ \emph{sets} that are combinations produced by taking subsets of the ten symbols $I_{10}$ three at a time. 
The choice of members of the sequence $N$ will be required to additionally satisfy:
\begin{equation*}
\begin{aligned}
	&|N \cap C_i| \le 1 \,\forall \, i \in I_{120} \\
\end{aligned}
\end{equation*}
This will guarantee that any two codewords differ as sets.
The problem, at first, appears to be amenable to bipartite matching or matroid intersection methods.
However, to guarantee that a latin square is produced,  each of the rows and columns must 
contain all the symbols as reflected in the conditions:
\begin{equation*}
\begin{aligned}
	& \forall \,r, s \in I_{10} \; | \{ (r\,s\,c_i)\; \ni\, i \in I_{10} \land \,(r\,s\,c_i)\,{\in}\,L\} |  = 1  \\
	& \forall \,s, c \in I_{10} \; | \{ (r_i\,s\,c)\; \ni\, i \in I_{10} \land \,(r_i\,s\,c)\,{\in}\,L\} |  = 1  \\
\end{aligned}
\end{equation*}
These conditions were translatable into a mixed integer linear program and were
presented to Sage~\wvcite{sage09} resulting in the desired permutation-free code within a few minutes.
The initial result did contain phonetic errors, until more constraints were added:
\begin{equation*}
\begin{aligned}
	& \forall \,r, c_i \in I_2^9 \; |\{ (1\,r\,c_i), (r\,0\,c_i) \}|\; \le 1 \; (\textit{left~phonetic}) \\
	&  \forall \,r_i, c \in I_2^9 \; |\{ (r_i\,1\,c), (r_i\,c\,0) \}|\; \le 1 \; (\textit{right~~phonetic})\\
\end{aligned}
\end{equation*}
The resulting code of Table~\wvref{Code47374800} is free of phonetic errors and completely handles
all error types mentioned in this paper except for triple errors.

\myTwoColSectionvspace
\section{Disjointness of Conjugate Codes}\label{almost}
Multiple codes that each have error detecting properties
and are pairwise disjoint can provide error detection when each code
is used for a single category of items while providing unique identifiers
for the entire collection. By reordering the components $(\,r\,s\,c\,)$ of each codeword, Table~\wvref{Code47374800} 
can  be interpreted in six ways to yield six codes which all have the same 
error correcting properties and are pairwise disjoint except for the 10
triple codewords replicated in each.
Again following \wvcite{McKay2007}, let $T = S_3$ the symmetric group on $3$ items
applied to permute the triplets used to represent codewords. 
The group actions are represented here as permutations of the symbols
$R$, $S$, and $C$ albeit with a movement of "S" to the middle position.
Thus, (RC) denotes a transpose of the table L, while a fixed point $x\in N$ for (SCR),
the other generator for~$T$,
would indicate a cyclic error.
The six resulting "conjugate" codes are then given by
\begin{equation*}
\begin{aligned}
	L_i = t_i(L),\; t_i\in T, \; \; \; i \in I_6
\end{aligned}
\end{equation*}
To see that this is permissible note that the determination of the third digit from the
other two is guaranteed by the latin square property to persist.
Actually, six phonetic error constraints were used in generation of the code
shown in  Table~\wvref{Code47374800} 
so that all six conjugate codes are free from phonetic errors.
Each constraint corresponds to the previously given left phonetic constraint
modified by application of a member of $T$ to its triplets.
Should there be data on errors to test these codes against, the properties of the six codes
espoused here will be unaffected by applying any permutation $P$ to the triplets of
the code $L$ of Table~\wvref{Code47374800} where $P$ has the form:
\begin{equation*}
\begin{aligned}
	&P( (r,s,c) ) = ( p(r)\,p(s)\,p(c) ) \\
	&p = p_{01} \circ p_{29}  \, \ni \, p_{01} \in S_{\{0, 1\}} \land\;p_{29} \in S_{\{2,\ldots, 9\}} \\
\end{aligned}
\end{equation*}
The properties of the codes are easily checked without regenerating them.
The author conjectures that non-isomorphic codes exist,
but leaves meaning of "isomorphic" \wvcite{McKay2007} to the reader.

\enlargethispage{\equalizelastpgcols}

\section{Remarks}
Evaluation of a code with respect to a potential use is a very complex problem.
There are doubtless circumstances where this code would fit nicely,
but it is presented here mainly as a design of combinatorial interest.
Given the table provided,  in many cases, 
no programming would be required as the uses for
length 3 check digit codes are generally static 
with one-time assignment of codewords. 
Dunning \wvcite{dunningMar2024} gives similar permutation-free codes
for other bases as well as a set of eight comparable length 3 decimal check digit codes
with only the single codeword \texttt{(999)} in common,
albeit with some cyclic and phonetic errors.

Consideration of the many well-designed codes from the extensive lists of references
given by Abdel-Ghaffer \wvcite{Abdel1998} and by Dunning \wvcite{dunningMar2024} is recommended 
when check digit codes with groupings of 4 digits or more are needed.
A remarkable sequence of codes for larger bases is given by Damm \wvcite{Damm2007}.
The text by Kirtland \wvcite{kirtland2001} provides a survey as well as references.

\iflatexml 
\section{Endnotes}

\begin{enumerate}
\item\label{foot1}\par{When referring to the original tables on pages 40-41 of \wvcite{Verhoeff1969}, 
the code shown inTable~\wvref{VerhoeffRegular}, ``resulting" from the block design, 
is erroneously interchanged with the ``interchanged" code (not shown) 
in which the codeword ``\texttt{109}" appeared as ``\texttt{100}".}

\item\label{foot2}\par{\copyright The author grants copyright permissions as specified in
Creative Commons License CC BY-SA 4.0.  
Attributions should reference the PDF version of this manuscript dated \today, 
which is preferred for printing and redistribution.
The author may be reached via e-mail:dunning@bgsu.edu at Larry A. Dunning, Professor Emeritus, 
Department of Computer Science, Bowling Green State University, Ohio 43403, USA.}
\end{enumerate}

\else 
\fi 


\iflatexml 
\else 
\fi 

\end{document}